\documentclass[preprint]{elsarticle}

\usepackage{lineno,hyperref}
\usepackage{amsmath,amssymb}
\usepackage{amsfonts}
\usepackage{bm}
\usepackage{mathrsfs}
\usepackage{natbib}
\biboptions{sort&compress}

\journal{Journal of \LaTeX\ Templates}

\newcommand{\di}{\mathrm{d}}
\newcommand{\modulo}[1]{\left |#1\right |}
\newcommand{\sqmodulo}[1]{{\modulo{#1}}^{2}}
\newcommand{\M}{{\cal{M}}}

\def\bit{\begin{itemize}}    \def\eit{\end{itemize}}
\def\ben{\begin{enumerate}}  \def\een{\end{enumerate}}
\def\bce{\begin{center}}    \def\ece{\end{center}}

\begin{document}

\begin{frontmatter}

\title{\textbf{Electroweak corrections to ${\bm e^+ \bm e^- \bm \to \bm \gamma \bm \gamma}$ as a luminosity process at FCC-ee} }

\author[mainaddressPV]{Carlo M. Carloni Calame}\corref{mycorrespondingauthor}
\cortext[mycorrespondingauthor]{Corresponding author}\ead{carlo.carloni.calame@pv.infn.it}
\author[mainaddressW]{Mauro Chiesa}
\author[secondaryaddressPV,mainaddressPV]{Guido Montagna}
\author[mainaddressPV]{Oreste~Nicrosini}
\author[mainaddressPV]{Fulvio Piccinini}

\address[mainaddressPV]{INFN, Sezione di Pavia, via A. Bassi 6, 27100 Pavia, Italy}
\address[mainaddressW]{Institut f\"ur Theoretische Physik und Astrophysik, Julius-Maximilians-Universit\"at W\"urzburg,
Emil-Hilb-Weg 22, D-97074 W\"urzburg, Germany}
\address[secondaryaddressPV]{Dipartimento di Fisica, Universit\`a di Pavia, via A. Bassi 6, 27100 Pavia, Italy}

\begin{abstract}
We consider large-angle two photon production 
in $e^+ e^-$ annihilation as a possible process 
to monitor the luminosity of a future  $e^+ e^-$ circular collider (FCC-ee). 
We review and assess the status of the theoretical accuracy by performing 
a detailed phenomenological study of next-to-leading order electroweak 
corrections and leading logarithmic QED contributions due to multiple photon radiation. 
We also estimate the impact of photonic and fermion-loop corrections at next-to-next-to-leading order
and the uncertainty induced by the hadronic contribution to the vacuum polarization. Possible perspectives to 
address the target theoretical accuracy are briefly discussed.
%\vskip 4pt
%{\it This article is registered under preprint number: ...}
\end{abstract}

\begin{keyword}
electron-positron colliders \sep luminosity \sep two photon production 
\sep QED \sep electroweak corrections \sep theoretical accuracy
\end{keyword}

\end{frontmatter}

\section{Introduction}

FCC-ee is a proposed high-luminosity $e^+ e^-$ circular collider under consideration at CERN 
as one of the accelerators for next-generation particle physics 
experiments~\cite{Bicer:2014,Abada:2019lih,Abada:2019zxq}. With a 
centre-of-mass energy (c.m.) between 90 and 365~GeV, it will provide the opportunity to test the 
Standard Model with unprecedented accuracy and perform indirect searches for New Physics 
through precision measurements. In particular, it could enable detailed investigations of the 
mechanism of electroweak symmetry breaking and high-precision studies of the properties of 
the $Z$, $W$, Higgs and top particles, as well as of the strong interaction.

The accomplishment of the above goals depends crucially on a number of 
critical factors, among which a precise knowledge of the collider luminosity. 
The ambitious FCC-ee target is a luminosity measurement with a total relative error 
of the order of $10^{-4}$ (or even better), not to spoil the expected statistical uncertainty of 
the main processes of interest. This precision exceeds that obtained at 
LEP and calls for a major effort by both the experimental and theoretical community. 

At FCC-ee, the standard luminosity process is expected to be $e^+ e^- \to e^+ e^-$ (Bhabha) 
scattering, as measured  by means of dedicated
calorimeters put in the very forward region close to the beams, likewise at LEP~\cite{ALEPH:2005ab}. 
In this respect, available precision calculations and related tools for 
small-angle Bhabha scattering have been reviewed in Ref.~\cite{Blondel:2018mad} and a possible path 
to 0.01\% theoretical luminosity accuracy has been outlined in Ref.~\cite{Jadach:2018jjo}.

However, also the process of large-angle photon-pair production, 
i.e. $e^+ e^- \to \gamma\gamma$, has been recently 
proposed as a possible alternative normalization process for the FCC-ee physics 
program~\cite{Janot:2015,Dam:2016,Carloni:2019}. 
It was already used to monitor the luminosity 
at $e^+ e^-$ flavor factories with c.m. energy at the 
GeV scale~\cite{Aloisio:2004bu,Dobbs:2007ab} and is still 
employed by BESIII Collaboration to cross-check the luminosity obtained by 
analyzing large-angle Bhabha scattering events~\cite{Ablikim:2018tdq}. Actually, 
$e^+ e^- \to \gamma\gamma$ turns out to be a particularly adequate normalization
process from a theoretical point of view. It is a purely QED process at leading order (LO)
at any energy, it receives QED corrections from the initial state only 
and does not contain at order $\alpha$ the contribution due to vacuum polarization 
(in particular, hadronic loops), which enters at next-to-next-to-leading order (NNLO) only. 
On the other hand, the process is affected by a large background due to large-angle 
Bhabha scattering,  which is huge on 
the $Z$ peak but more manageable at higher energies.

In spite of the above limitation, the possibility of using two photon production as 
a luminosity process at FCC-ee is an interesting option to be pursued. 
Contrarily to Bhabha scattering, that received a lot of attention over the past decades, 
there is a rather poor theoretical literature about $e^+ e^- \to \gamma\gamma$ annihilation and the most recent 
phenomenological results refer to $e^+ e^-$ colliders of 
moderate energies~\cite{Arbuzov:1997pj,Balossini:2008xr,Actis:2010gg,Eidelman:2010fu}. 
Also the available Monte Carlo (MC) generators~\cite{Eidelman:2010fu,Balossini:2008xr}, which are necessary for experimental simulations
and feasibility studies, are tailored for low-energy accelerators and need to be improved
for the high-energy, high-precision requirements of FCC-ee.

With these motivations in mind, we take a first step towards a critical assessment of 
the current status of the theoretical accuracy for large-angle photon-pair production at FCC-ee energies. 
 For this purpose, we make use of the MC program 
\textsc{BabaYaga@nlo}~\cite{CarloniCalame:2000pz,CarloniCalame:2001ny,CarloniCalame:2003yt,Balossini:2006wc,Balossini:2008xr}\footnote{
It was extensively used and is still a reference code for luminosity measurements at flavor factories.}
and improve it 
by including purely weak corrections due to heavy boson exchange. We examine in detail 
the effects of next-to-leading (NLO) electroweak corrections and higher-order QED contributions for both 
integrated and differential cross sections and according to different event selection criteria.
We also explore the r\^ole played by photonic and fermion-loop corrections at NNLO and the uncertainty
driven by the hadronic contribution to the vacuum polarization. QED corrections to 
$e^+ e^- \to \gamma\gamma$ at order $\alpha$ were previously 
computed in Refs.~\cite{Berends:1973tm,Eidelman:1978rw,Berends:1980px}
and NLO electroweak corrections were obtained in Refs.~\cite{CapdequiPeyranere:1978ce,Bohm:1986dn,Fujimoto:1986xb}. 
A generator based on Ref.~\cite{Berends:1980px} was used at LEP for the 
analysis of photon-pair production at energies above the $Z$ mass~\cite{Alcaraz:2006mx}. 
It is worth noting that the full set of electroweak corrections is unavailable 
in any modern MC generator
but the version of \textsc{BabaYaga@nlo} developed for this study, which also provides the exponentiation of QED leading logarithmic
(LL) contributions due to multiple photon radiation.

With respect to our contribution to the workshop proceedings as in
Ref.~\cite{Blondel:2019vdq}, the present paper provides theoretical
details and further numerical results relevant for two photon production as a
luminosity process at FCC-ee.

The structure of the article is as follows. In Sect.~2 we sketch the theoretical 
formulation of the QED radiation inherent to \textsc{BabaYaga@nlo} and the computation 
of one-loop weak corrections. The results of our numerical study are shown in Sect.~3.
In Sect.~4 we draw the main conclusions of our analysis and discuss possible ways to 
achieve the target theoretical accuracy of FCC-ee.

\section{Electroweak corrections}

The photonic and weak corrections to $e^+ e^- \to \gamma\gamma$ form two 
gauge-invariant subsets and can be treated separately.

According to the theoretical formulation implemented in \textsc{BabaYaga@nlo}, the photonic corrections
are computed by using a fully-exclusive QED Parton Shower (PS) matched to QED 
contributions at NLO.
The master formula for the cross section calculation reads as follows:
\begin{equation}
\di\sigma = F_{\textrm{SV}}\,\Pi^2 \left(Q^{2},\epsilon\right)\sum_{n=0}^{\infty}\dfrac{1}{n!}\left(\prod_{i=0}^{n}F_{\textrm{H},i}\right)\modulo{\M_{n,\textrm{LL}}}^{2}\di \Phi_{n}
\label{eq:1}
\end{equation}
In Eq.~(\ref{eq:1}), $\Pi\left(Q^{2},\epsilon\right)$ is the Sudakov form factor, which accounts for the 
exponentiation of LL  contributions due to soft and virtual corrections, $Q^2$ and $\epsilon$ 
being the hard scale of the process and a soft-hard photon separator, respectively. We set $Q^2 = s$, where $s$ is
the squared c.m. energy, in such a way that the big collinear logarithms %of the kind
$L = \ln (s / m_e^2)$ due to initial-state radiation are resummed to all orders. 
$\sqmodulo{\M_{n,\textrm{LL}}}$ is the squared matrix element describing the emission in LL 
approximation of $n$ hard photons (i.e. with energy larger than 
$\epsilon$) on top of the two LO ones. The 
factor $\di\Phi_{n}$ is the exact phase space element of the process $2\gamma$
 plus $n$ additional radiated photons.

In Eq.~(\ref{eq:1}) the matching of the above PS ingredients with the NLO QED corrections is realized 
by the finite-order factors $F_{\textrm{SV}}$ and $F_{\textrm{H}}$, whose definitions can be found in 
Refs.~\cite{Balossini:2008xr,Balossini:2006wc}. 
They are infrared/collinear safe correction factors that account for 
those $O(\alpha)$ non-logarithmic terms entering the NLO calculation and absent in the PS approach. 
For the soft+virtual factor $F_{\textrm{SV}}$ we use the results of Ref.~\cite{Berends:1980px}, while the matrix element 
of the radiative process $e^+ e^- \to \gamma\gamma\gamma$, which is needed for the calculation of the hard bremsstrahlung 
factor $F_{\textrm{H}}$, is computed by means of the symbolic manipulation 
program \textsc{Form}~\cite{Kuipers:2012rf,Ruijl:2017dtg}.

By construction, the matching procedure as in Eq.~(\ref{eq:1}) is such that the $O(\alpha)$ expansion 
of Eq.~(\ref{eq:1}) reproduces the NLO cross section and exponentiation of LL contribution is preserved 
as in a pure PS algorithm. Moreover, as a by-product of its factorized structure, the bulk of the photonic 
sub-leading contributions at NNLO, i.e those of the order of $\alpha^2 L$, is automatically included 
by means of terms of the type $F_{{\textrm{SV}}~|~{\textrm{H}}}~\otimes$~LL corrections~\cite{Montagna:1996gw}.

To meet the high-energy, high-precision requirements of FCC-ee, we improved the theoretical content of
\textsc{Babayaga@nlo} by calculating the one-loop weak corrections due to $W$, $Z$ and Higgs exchange. 
We computed them using the computer program \textsc{Recola}~\cite{Actis:2012qn,Actis:2016mpe}, 
which internally adopts the \textsc{Collier}~\cite{Denner:2016kdg} 
library for the evaluation of 
one-loop scalar and tensor integrals. The calculation has been performed in the on-shell 
renormalization scheme, with complex mass values for the heavy 
boson masses~\cite{Denner:1999gp,Denner:2005fg,Denner:2006ic}.

The fermion-loop corrections to the photon self energy, which are needed for the estimate 
of the two loop contributions addressed 
in the next Section\footnote{At NNLO, there is a contribution to fermion-loop 
corrections arising from light-by-light scattering, which is not considered in the 
present study and whose evaluation is left to future work.}, 
are taken into account using the following
expression for the vacuum polarization correction
\begin{equation}
\Delta\alpha(s) \, = \, \Delta\alpha_{\rm lep} (s) + \Delta\alpha_{\rm had} (s) + \Delta\alpha_{\rm top} (s)
\label{eq:2}
\end{equation}
For the leptonic correction $\Delta\alpha_{\rm lep} (s)$ and the top-quark contribution $\Delta\alpha_{\rm top} (s)$
we use the well known results in one-loop approximation. The hadronic (light-quark) correction is accounted for
according to a dispersive approach based on time-like data for the process $e^+ e^- \to$~hadrons, as 
implemented in the latest 
version of the {\tt hadr5n16.f} routine~\cite{Jegerlehner:2017zsb}\footnote{Available at {\tt http://www-com.physik.hu-berlin.de/\~{}fjeger/software.html}}.

\section{Numerical results}

For the presentation of the results of our numerical study, we use the following set of 
input parameters:
\begin{eqnarray}
&& \alpha(0) = 1 / 137.03598950000034580626 \nonumber\\
&& M_Z   =  91.15348 \, {\rm GeV}    \quad   ~~~     \Gamma_Z   =  2.49427 \, {\rm GeV} \nonumber\\
&& M_W   =  80.35797  \, {\rm GeV}   \quad    ~~   \Gamma_W   =  2.08430 \,  {\rm GeV} \nonumber\\
&& M_H   =  125 \, {\rm GeV}    \nonumber\\ 
&& m_e   = 0.51099 \, {\rm MeV}    \qquad  ~~  m_\mu  = 0.10566 \, {\rm GeV} \nonumber\\
&& m_{\tau} =  1.777 \, {\rm GeV}   \qquad ~~~~~ m_{\rm top} =  173.2 \, {\rm GeV}
\end{eqnarray}

We consider four c.m. energy values, which are representative of the expected FCC-ee operation 
program ($Z$-pole, $WW$, $ZH$ and $t \bar{t}$ tresholds)
\begin{equation}
\sqrt{s}=91,\ 160,\ 240,\ 365 \ {\rm GeV}
\end{equation}

For those c.m. energies, the collinear logarithm $L$ varies in the range between 24 and 27, 
and the QED expansion parameter $\beta = 2 \ \alpha / \pi (L-1) $ is of the order of 0.1.

To study the dependence of the QED corrections on the applied cuts, we consider two different 
simulation setup
\bit
\item[] {[a]} Full phase space, i.e. no cuts
\item[] {[b]} Acceptance cuts, i.e. at least two photons with: $20^\circ<\theta_\gamma<160^\circ$ and \\ $E_\gamma\ge 0.25\times\sqrt{s} $
\eit

In Tab.~\ref{Tab:Tab1} and Tab.~\ref{Tab:Tab2}, we examine the impact of the QED radiative 
corrections on the integrated cross sections, when considering setup [a] (Tab.~\ref{Tab:Tab1})
and setup~[b] (Tab.~\ref{Tab:Tab2}).
\begin{table}[hbtp]
\bce
\begin{tabular}[c]{r||c|c|c}
$\sqrt{s}$ (GeV) & LO (pb) & NLO (pb) & w h.o. (pb)\\[1mm]
\hline
      &          &                    &   \\[-3mm]
$91$  & $364.68$ & $447.27$ $[+23\%]$ & $445.6(9)$  $[-0.46\%]$ \\ 
$160$ & $123.71$ & $154.37$ $[+25\%]$ & $153.2(2)$  $[-0.95\%]$ \\ 
$240$ & $56.816$ & $71.809$ $[+26\%]$ & $71.07(6)$  $[-1.30\%]$ \\ 
$365$ & $25.385$ & $32.515$ $[+28\%]$ & $32.09(2)$  $[-1.67\%]$ \\
\hline
\end{tabular}
\caption{The two photon production cross section at LO, NLO QED and according to Eq.~(\ref{eq:1}),  
for four FCC-ee c.m. energies and according to setup [a]. The numbers in parenthesis are the 
relative contributions of NLO and higher-order LL corrections.}
\label{Tab:Tab1}
\ece
\end{table}

\begin{table}[hbtp]
\bce
\begin{tabular}[c]{r||c|c|c|c}
$\sqrt{s}$ (GeV) & LO (pb) & NLO (pb) & w h.o. (pb) & Bhabha LO (pb)\\[1mm]
\hline
      &          &                       &                          &\\[-3mm]
$91$  & $39.821$ & $41.043$  $[+3.07\%]$ & $40.870(4)$    $[-0.43\%]$ & $2625.9$\\   
$160$ & $12.881$ & $13.291$  $[+3.18\%]$ & $13.228(1)$    $[-0.49\%]$ & $259.98$\\  
$240$ & $5.7250$ & $5.9120$  $[+3.27\%]$ & $5.8812(6)$   $[-0.54\%]$ & $115.77$ \\   
$365$ & $2.4752$ & $2.5581$  $[+3.35\%]$ & $2.5438(3)$    $[-0.58\%]$ & $50.373$  \\ 
\hline
\end{tabular}
\caption{The same as in Tab.~\ref{Tab:Tab1} according to the cuts of setup [b]. In the last column, 
the LO large-angle Bhabha cross section in the same setup is shown for the sake of comparison.}
\label{Tab:Tab2}
\ece
\end{table}
The photon-pair production cross section is shown 
according to different accuracy levels, i.e. at LO, NLO QED and according to Eq.~(\ref{eq:1}), 
that includes the LL contributions due to multiple photon radiation. The numbers in parenthesis are the 
relative contributions due to NLO and higher-order LL corrections to LO, respectively. 
It can be noticed that the NLO corrections are particularly relevant in the inclusive setup [a], 
varying in the range between 20\% and 30\%, whereas they get largely reduced, to a few percent level, 
in setup [b]. The same trend is observed for the higher-order contributions, which range 
between 0.5\% and 2\% in setup [a], while they always amount to about five per mille in setup [b], 
independently of the c.m. energy. The difference between the results of the two configurations can be 
ascribed to the presence in setup [a] of logarithmic enhancements due to radiation of photons 
emitted collinear to the beams, that are largely removed by the angular cuts of setup [b]. This suggests that the overall 
contribution of QED corrections can be made sufficiently small by means of appropriate 
event selections, 
keeping at the same time an acceptable statistical uncertainty thanks to the high FCC-ee luminosity.

The higher-order effects discussed above are dominated by $O(\alpha^2 L^2)$ 
LL corrections but also include sub-leading contributions beyond NLO, 
among which $O(\alpha^2 L)$ photonic corrections. As already remarked, they 
originate as a by-product of the factorized form of Eq.~(\ref{eq:1}) and their contribution
can be estimated according to the following chain formula~\cite{Balossini:2008xr,Balossini:2006wc}
\begin{equation}
\delta_{\alpha^2 L} \, = \, \dfrac{\sigma - \sigma_{\textrm{NLO}} - 
\sigma_{\textrm{exp}}^{\textrm{PS}}+\sigma_{\alpha}^{\textrm{PS}}}{{\sigma}_{\textrm{LO}} }
\label{eq:5}
\end{equation}

 \begin{figure}[t]
\centering
\includegraphics[width=9.cm]{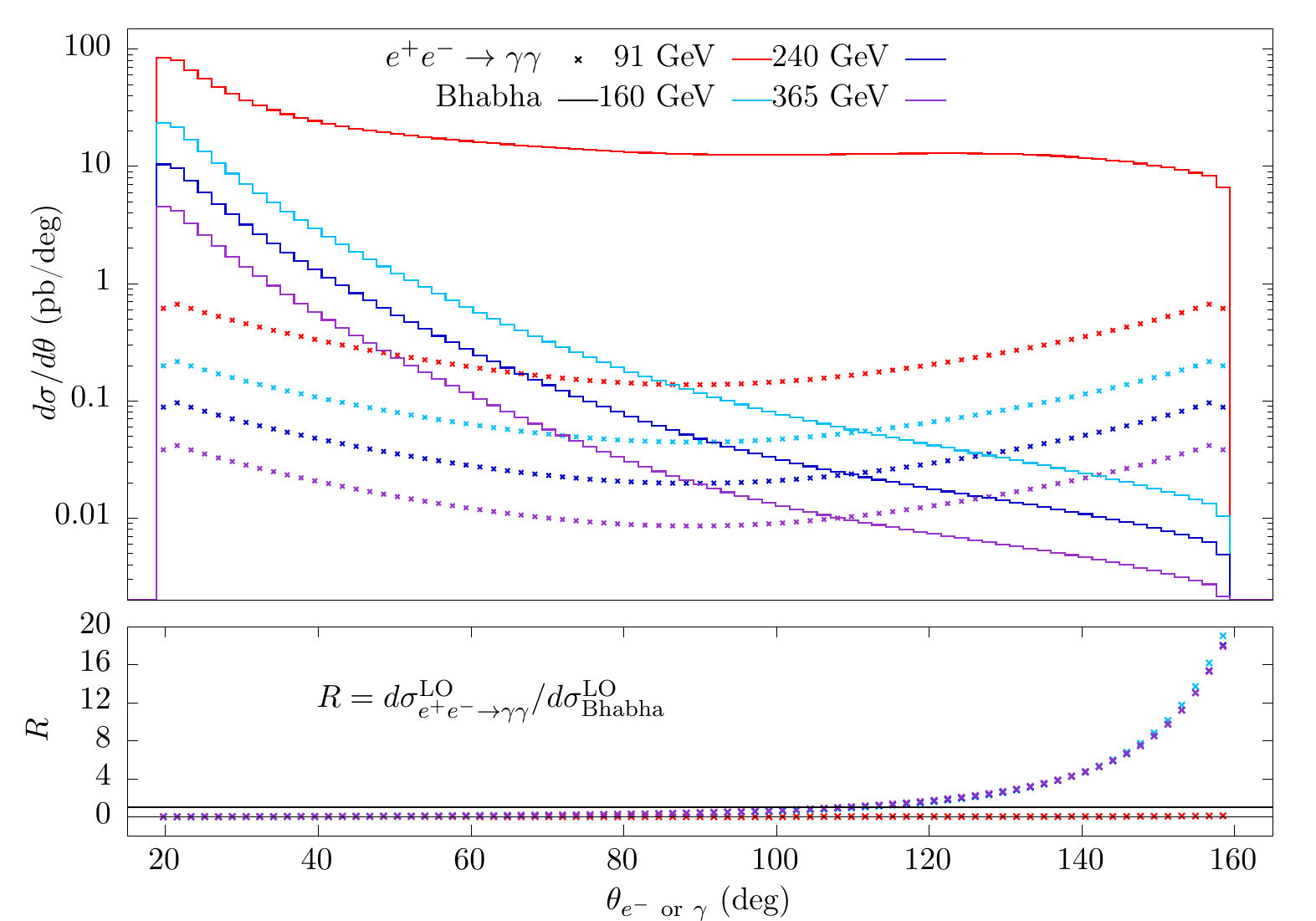}
\caption{Upper panel: the LO angular distributions of $e^+ e^- \to \gamma\gamma$ and 
large-angle Bhabha scattering, for four FCC-ee c.m. energies and according to setup [b]. 
Lower panel: ratio of the differential cross sections of the two processes.}
\label{Fig:Fig4}
\end{figure}

In Eq.~(\ref{eq:5}), $\sigma$ is the full factorized cross section as in Eq.~(\ref{eq:1}), $\sigma_{\textrm{NLO}}$ 
the exact NLO cross section, $\sigma_{\textrm{exp}}^{\textrm{PS}}$ and 
$\sigma_{\alpha}^{\textrm{PS}}$ are the pure PS cross sections with all-order and $O(\alpha)$ 
corrections, respectively. An estimate of $O(\alpha^2 L)$ contributions is relevant because it allows to probe the 
size of the most relevant higher-order corrections beyond the LL approximation~\cite{Jadach:2018jjo,Montagna:1996gw}.
By using Eq.~(\ref{eq:5}), we studied the impact of $O(\alpha^2 L)$ photonic corrections as a 
function of the selection criteria, also considering an additional acollinearity cut of $10^\circ$ that tends to single out elastic events.
We obtained that $\delta_{\alpha^2 L}$ varies in setup [b] in the few per mille range as 
a function of $\sqrt{s}$ and gets reduced below 0.1\% when adding the acollinearity cut to 
the conditions of setup~[b].

For the sake of comparison, we also show in Tab.~\ref{Tab:Tab2} the LO cross section of the large-angle 
Bhabha process, which turns out to be the main background to the two photon signature. 
The Bhabha scattering cross section is evaluated using the same cuts of event selection [b] applied to
$e^+ e^- \to \gamma\gamma$. As can be seen, Bhabha scattering provides at the $Z$ resonance an 
overwhelming background ($\sigma_{\rm Bhabha} \simeq 66\times\sigma_{\gamma\gamma}$), 
which remains sizeable but less important for the other energy points ($\sigma_{\rm Bhabha} \simeq 20\times\sigma_{\gamma\gamma}$).
\begin{figure}[hbtp]
\centering
\includegraphics[width=9.cm]{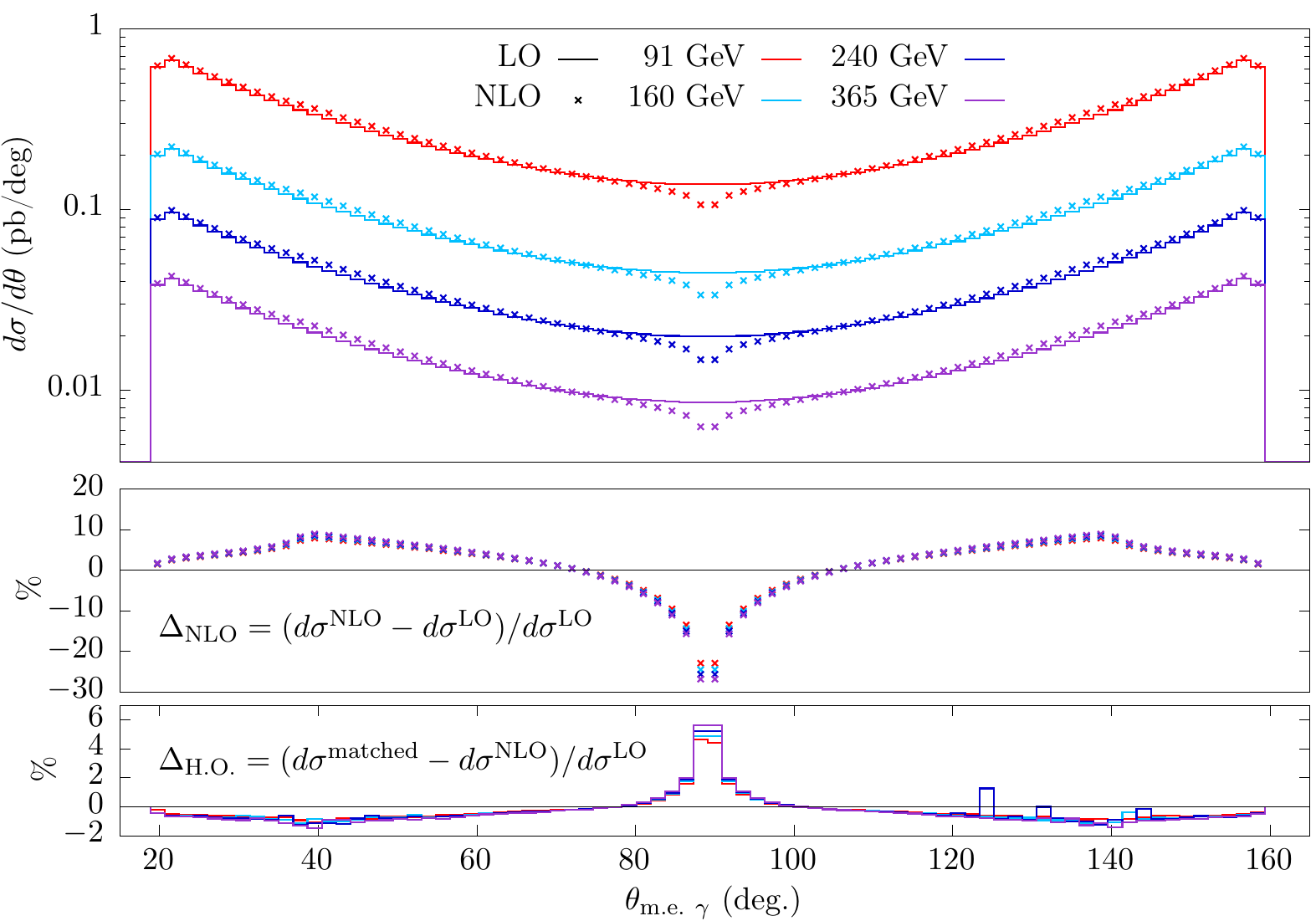}
\caption{Upper panel: the angular distribution of the most energetic photon, 
for four FCC-ee c.m. energies and according to setup [b]. Lower panel: 
relative contributions of NLO and higher-order LL corrections.}
\label{Fig:Fig5}
\end{figure}
To get further insight on the interplay between the two processes, we show in Fig.~\ref{Fig:Fig4} 
a comparison between the LO angular distributions of $e^+ e^- \to \gamma\gamma$ and 
large-angle Bhabha scattering. The distributions refer to the four c.m. energies and to the cuts 
of setup [b] for both processes. As usual, the scattering angles are defined with respect 
to the direction of the incoming electron. In the upper panel of Fig.~\ref{Fig:Fig4}, 
one can notice the 
strongly asymmetric behavior of Bhabha scattering, which is dominated by $t$-channel
 photon exchange, in comparison with the symmetric shape of $e^+ e^- \to \gamma\gamma$. 
 As can be seen, the cross section of two photon production is significantly larger than that of Bhabha scattering 
in the backward $\theta_{e^-}$ hemisphere (large electron scattering angles) for all energies, except at the $Z$ resonance. 
Obviously, the same holds for the ratio of the two distributions as a function of $\theta_{e^+}$ in the 
forward $\theta_{e^-}$ emisphere, i.e. at small positron scattering angles. This simple analysis seems to
suggest that, at least well above the $Z$ pole, the angular distributions can provide a handle to control the Bhabha background, 
provided a sufficiently good discrimination of the electron/positron charge is performed in the experiment.

A representative example of the effects due to QED corrections on the differential cross sections is given in Fig.~\ref{Fig:Fig5}, 
which shows the angular distribution of the most energetic photon for the four energy points, according 
to setup [b]. As can be noticed,
the NLO corrections are particularly important in the central region, where they are large and negative, reaching the 
20-30\% level, as mainly due to soft-photon radiation. This effect is partially compensated by higher-order
corrections, that amount to some percents in the same region. 

\begin{figure}[hbtp]
\centering
\includegraphics[width=9.cm]{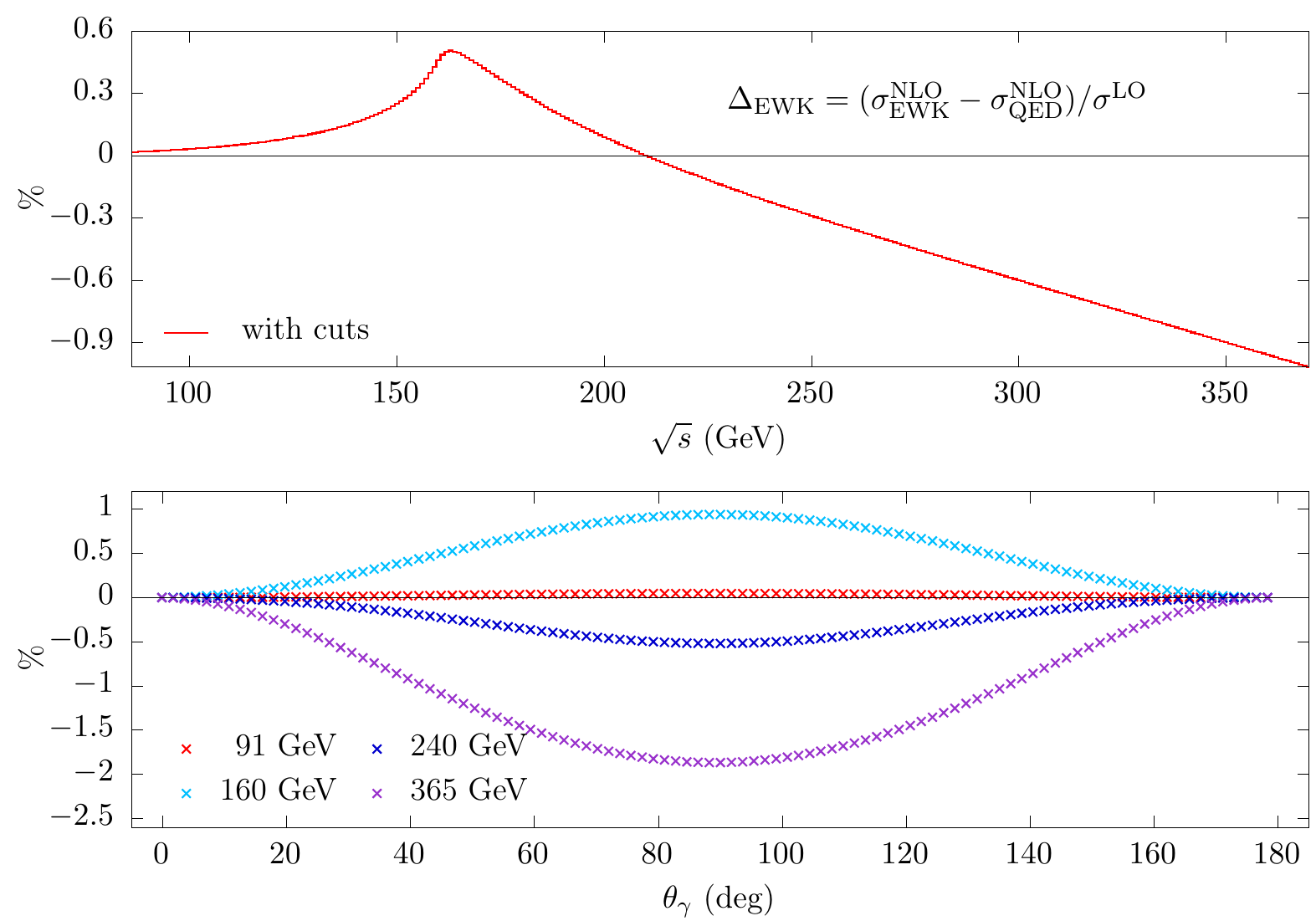}
\caption{Upper plot: relative contribution of the NLO weak corrections to the 
integrated cross section according to setup [b], as a function of the c.m. energy. Lower plot: the same as in the upper
plot for the photon angular distribution, at four FCC-ee c.m. energies.}
\label{Fig:Fig6}
\end{figure}

In Fig.~\ref{Fig:Fig6} we show the contribution of weak corrections to the integrated cross section as a function of the c.m. energy
in the case of setup [b] and to the photon angular distribution at the four canonical energy points.
As expected, the correction to the integrated cross section is of increasing importance as the energy increases, 
varying from a few per mille to one per cent. It amounts to about 0.5\% around the $W$-pair production threshold, 
it passes through zero around the $ZH$ threshold and becomes more and more negative from $ZH$ to
the $t \bar{t}$ production thresholds. 
Concerning the angular distribution, the contribution of 
weak corrections is practically negligible at the $Z$ resonance, at the per cent level for the other energy points and
more pronounced in the central region for any energy, where it is 
of the same order as higher-order QED contributions at high 
energies.

Our estimate of the fermion-loop correction to the integrated cross section is given in Tab.~\ref{Tab:Tab3},  
using for definitiveness the setup [b] that includes acceptance cuts. The numerical results of Tab.~\ref{Tab:Tab3} 
are obtained by factorization of the NLO photonic correction with the vacuum polarization contribution 
according to the following formula
\begin{equation}
\sigma^\text{NNLO}_{\Delta\alpha} \, \pm \, \delta\sigma_{\rm had} \, \simeq  \left(\sigma^\text{NLO}_\text{QED}-\sigma^\text{LO}\right) 
 \times\left[\Delta\alpha (s) \, \pm \, \delta\Delta\alpha_{\rm had} \right]
\label{eq:6}
\end{equation}
where $\delta\Delta\alpha_{\rm had}$ is the data-driven uncertainty due to the hadronic contribution to $\Delta\alpha$, as 
returned by the  {\tt hadr5n16.f} routine. The factorized approach as in Eq.~(\ref{eq:6}) 
gives rise to corrections dominated by $O(\alpha^2 L^2)$ contributions and was proved in Ref.~\cite{CarloniCalame:2011zq} 
to be an excellent approximation of the perturbative result based on an exact 
NNLO calculation.
\begin{table}[hbtp]
\bce
\begin{tabular}[c]{r||c|c|c}
  $\sqrt{s}$ (GeV) & $\sigma^\text{NNLO}_{\Delta\alpha \, {\rm lep+top}}/\sigma_{LO}$ &  
  $\sigma^\text{NNLO}_{\Delta\alpha \, {\rm had}}/\sigma_{LO}$& $\delta\sigma_{\rm had}/\sigma_{LO}$\\[1mm]
  \hline
  & & & \\[-3mm]
  $91$ &$ 0.096\%$ & $0.085\%$ & $3.7\cdot10^{-6}$\\
 $160$ &$0.108\%$ & $0.098\%$ & $3.8\cdot10^{-6}$\\
 $240$ &$0.115\%$ & $0.108\%$ & $3.9\cdot10^{-6}$\\
 $365$ &$0.119\%$ & $0.120\%$ & $4.0\cdot10^{-6}$\\
 \hline
\end{tabular}
\caption{Relative contribution of the NNLO leptonic(+top) and hadronic vacuum polarization 
correction to the cross section in setup [b] and for four FCC-ee c.m. energies. In the last column, 
the uncertainty due to the hadronic contribution is shown.}
\label{Tab:Tab3}
\ece
\end{table}
As can be seen, the vacuum polarization correction due to both leptonic(+top) and hadronic loops amounts 
to about 0.1\% at all c.m. energies. In particular, we checked that top-quark contribution is always completely negligible. 
Note that the parametric uncertainty induced by the hadronic contribution to $\Delta\alpha$ is much 
smaller than the target accuracy and therefore is not a limiting factor for the theoretical predictions 
for $e^+ e^- \to \gamma\gamma$. This is a strength of two photon production and 
is in contrast to small-angle Bhabha scattering, where the same
uncertainty presently contributes at the $10^{-4}$ level~\cite{Jadach:2018jjo}. 
It must be 
also noticed that a sound assessment of this class of corrections requires 
an explicit two-loop computation, as well as 
the combination of loop effects with the same-order contribution 
of real pair emission, which partially cancels the fermion-loop correction, as shown in past precision calculations for 
Bhabha scattering~\cite{Jadach:1993wk,Arbuzov:1995qd,Montagna:1998vb,CarloniCalame:2011zq}.
          
\section{Conclusions}

We have studied large-angle two photon production in $e^+ e^-$ annihilation as a possible process 
to measure the luminosity at FCC-ee experiments. We have assessed the 
status of the theoretical accuracy by performing a thorough phenomenological study of the radiative 
corrections to $e^+ e^- \to \gamma\gamma$ annihilation for all the relevant c.m. energies. 
To that purpose, we have upgraded the theoretical content of the code \textsc{BabaYaga@nlo}, that 
includes exact NLO QED corrections matched to PS, by computing the weak corrections due to the presence 
of heavy bosons in the internal loops.

We have shown that in a realistic setup including acceptance cuts the NLO QED corrections are fairly small,
being at the level of a few percents for all the relevant c.m. energies. In the same conditions, the effects
due to multiple photon emission, which are dominated by $O(\alpha^2 L^2)$ contributions, 
amount to about 0.5\%. The one-loop weak 
corrections to the integrated cross section and angular distribution are below 1\% and 
at a few percent level, respectively.
As a whole, these results point out that NLO electroweak and higher-order QED corrections to 
photon-pair production at FCC-ee are moderate but strictly necessary 
for precision luminosity monitoring.

We have also probed the size of some radiative corrections entering at NNLO accuracy. 
We have shown that sub-leading $O(\alpha^2 L)$ photonic contributions vary from 0.01\% to a 
few 0.1\%, depending on the applied cuts and the considered c.m. energy.
We have also estimated the vacuum polarization correction due to leptonic and hadronic loops, 
to conclude that it provides a contribution at the
one per mille level for all the c.m. energies of interest. 
Moreover, we have shown that the uncertainty induced by the hadronic loops 
is below $10^{-5}$ and therefore does not provide a limitation for the theoretical predictions, 
contrarily to small-angle Bhabha scattering. Note that at NNLO accuracy fermion-loop corrections also involve 
the contribution from light-by-light scattering. The evaluation of this effect and of its uncertainty is left
to future work.

As far as QED corrections are concerned, the accuracy of the present calculation can be estimated to be 
at the 0.1\% level or slightly better~\cite{Balossini:2006wc,Balossini:2008xr,Actis:2010gg,CarloniCalame:2011zq}.
However, the above conclusions about photonic and vacuum polarization corrections at NNLO suggest that the theoretical formulation 
implemented in our code, if supplemented by the contribution of $O(\alpha^2)$ fermion-loop and real pair corrections, should 
be sufficient to get close to an accuracy at the $10^{-4}$ level. Previous results about Bhabha scattering as luminosity process
at LEP~\cite{Montagna:1998vb,Jadach:2018jjo} 
and flavor factories~\cite{Balossini:2006wc,Actis:2010gg,CarloniCalame:2011zq} support this expectation. 
Beyond a 0.01\% accuracy, a full calculation of NNLO QED corrections and, eventually, of two-loop
weak contributions will be ultimately needed to reach the challenging frontier of 
the 10~ppm theoretical accuracy. 

\vskip 12pt \noindent
{\bf Acknowledgements}
\vskip 8pt\noindent
We wish to thank A.~Blondel, M.~Dam, J.~Gluza, S.~Jadach, P.~Janot and R.~Tenchini for interest in our work and useful discussions
in the context of FCC-ee Physics Workshops.

\bibliography{ggb}

\begin{thebibliography}{10}
\expandafter\ifx\csname url\endcsname\relax
  \def\url#1{\texttt{#1}}\fi
\expandafter\ifx\csname urlprefix\endcsname\relax\def\urlprefix{URL }\fi
\expandafter\ifx\csname href\endcsname\relax
  \def\href#1#2{#2} \def\path#1{#1}\fi

\bibitem{Bicer:2014}
M.~Bicer, et~al., {First Look at the Physics Case of TLEP}, JHEP 01 (2014) 164.
\newblock \href {http://arxiv.org/abs/1308.6176} {\path{arXiv:1308.6176}}.

\bibitem{Abada:2019lih}
A.~Abada, et~al., {FCC Physics Opportunities}, Eur. Phys. J. C79~(6) (2019)
  474.
\newblock \href {http://dx.doi.org/10.1140/epjc/s10052-019-6904-3}
  {\path{doi:10.1140/epjc/s10052-019-6904-3}}.

\bibitem{Abada:2019zxq}
A.~Abada, et~al., {FCC-ee: The Lepton Collider}, Eur. Phys. J. ST228~(2) (2019)
  261.
\newblock \href {http://dx.doi.org/10.1140/epjst/e2019-900045-4}
  {\path{doi:10.1140/epjst/e2019-900045-4}}.

\bibitem{ALEPH:2005ab}
S.~Schael, et~al., {Precision electroweak measurements on the $Z$ resonance},
  Phys. Rept. 427 (2006) 257--454.
\newblock \href {http://arxiv.org/abs/hep-ex/0509008}
  {\path{arXiv:hep-ex/0509008}}, \href
  {http://dx.doi.org/10.1016/j.physrep.2005.12.006}
  {\path{doi:10.1016/j.physrep.2005.12.006}}.

\bibitem{Blondel:2018mad}
A.~Blondel, et~al., {Standard Model Theory for the FCC-ee: The Tera-Z}, in:
  {Mini Workshop on Precision EW and QCD Calculations for the FCC Studies :
  Methods and Techniques CERN, Geneva, Switzerland, January 12-13, 2018}, 2018.
\newblock \href {http://arxiv.org/abs/1809.01830} {\path{arXiv:1809.01830}}.

\bibitem{Jadach:2018jjo}
S.~Jadach, W.~Placzek, M.~Skrzypek, B.~F.~L. Ward, S.~A. Yost, {The path to
  0.01\% theoretical luminosity precision for the FCC-ee}, Phys. Lett. B790
  (2019) 314--321.
\newblock \href {http://arxiv.org/abs/1812.01004} {\path{arXiv:1812.01004}},
  \href {http://dx.doi.org/10.1016/j.physletb.2019.01.012}
  {\path{doi:10.1016/j.physletb.2019.01.012}}.

\bibitem{Janot:2015}
P.~Janot, {Determination of $\alpha_{\rm QED} (M_Z)$ @ FCC-ee}, Talk given at
  FCC-ee Physics meeting, June 2015, CERN (2015).

\bibitem{Dam:2016}
M.~Dam, {Luminosity measurements at FCC-ee}, Talk given at FCC week, April
  2016, Rome (2016).

\bibitem{Carloni:2019}
C.~M. Carloni~Calame, {$e^+ e^- \to \gamma\gamma$ at large angle for FCC-ee
  luminometry}, Talk given at $11^{\rm th}$ FCC-ee workshop: Theory \&
  Experiments, January 2019, CERN (2019).

\bibitem{Aloisio:2004bu}
A.~Aloisio, et~al., {Measurement of $\sigma(e^+e^- \to \pi^+ \pi^- \gamma$) and
  extraction of $\sigma(e^+e^- \to \pi^+ \pi^-$) below 1-GeV with the KLOE
  detector}, Phys. Lett. B606 (2005) 12--24.
\newblock \href {http://arxiv.org/abs/hep-ex/0407048}
  {\path{arXiv:hep-ex/0407048}}, \href
  {http://dx.doi.org/10.1016/j.physletb.2004.11.068}
  {\path{doi:10.1016/j.physletb.2004.11.068}}.

\bibitem{Dobbs:2007ab}
S.~Dobbs, et~al., {Measurement of absolute hadronic branching fractions of D
  mesons and $e^+ e^- \to D \bar{D}$ cross-sections at the $\psi(3770)$}, Phys.
  Rev. D76 (2007) 112001.
\newblock \href {http://arxiv.org/abs/0709.3783} {\path{arXiv:0709.3783}},
  \href {http://dx.doi.org/10.1103/PhysRevD.76.112001}
  {\path{doi:10.1103/PhysRevD.76.112001}}.

\bibitem{Ablikim:2018tdq}
M.~Ablikim, et~al., {Measurement of the integrated luminosities of
  cross-section scan data samples around the $\psi(3770)$ mass region}, Chin.
  Phys. C42~(6) (2018) 063001.
\newblock \href {http://arxiv.org/abs/1803.03802} {\path{arXiv:1803.03802}},
  \href {http://dx.doi.org/10.1088/1674-1137/42/6/063001}
  {\path{doi:10.1088/1674-1137/42/6/063001}}.

\bibitem{Arbuzov:1997pj}
A.~B. Arbuzov, G.~V. Fedotovich, E.~A. Kuraev, N.~P. Merenkov, V.~D. Rushai,
  L.~Trentadue, {Large angle QED processes at $e^+ e^-$ colliders at energies
  below 3-GeV}, JHEP 10 (1997) 001.
\newblock \href {http://arxiv.org/abs/hep-ph/9702262}
  {\path{arXiv:hep-ph/9702262}}, \href
  {http://dx.doi.org/10.1088/1126-6708/1997/10/001}
  {\path{doi:10.1088/1126-6708/1997/10/001}}.

\bibitem{Balossini:2008xr}
G.~Balossini, C.~Bignamini, C.~M. Carloni~Calame, G.~Montagna, O.~Nicrosini,
  F.~Piccinini, {Photon pair production at flavour factories with per mille
  accuracy}, Phys. Lett. B663 (2008) 209--213.
\newblock \href {http://arxiv.org/abs/0801.3360} {\path{arXiv:0801.3360}},
  \href {http://dx.doi.org/10.1016/j.physletb.2008.04.007}
  {\path{doi:10.1016/j.physletb.2008.04.007}}.

\bibitem{Actis:2010gg}
S.~Actis, et~al., {Quest for precision in hadronic cross sections at low
  energy: Monte Carlo tools vs. experimental data}, Eur. Phys. J. C66 (2010)
  585--686.
\newblock \href {http://arxiv.org/abs/0912.0749} {\path{arXiv:0912.0749}},
  \href {http://dx.doi.org/10.1140/epjc/s10052-010-1251-4}
  {\path{doi:10.1140/epjc/s10052-010-1251-4}}.

\bibitem{Eidelman:2010fu}
S.~I. Eidelman, G.~V. Fedotovich, E.~A. Kuraev, A.~L. Sibidanov, {Monte-Carlo
  Generator Photon Jets for the process $e^+e^- \to \gamma \gamma$}, Eur. Phys.
  J. C71 (2011) 1597.
\newblock \href {http://arxiv.org/abs/1009.3390} {\path{arXiv:1009.3390}},
  \href {http://dx.doi.org/10.1140/epjc/s10052-011-1597-2}
  {\path{doi:10.1140/epjc/s10052-011-1597-2}}.

\bibitem{CarloniCalame:2000pz}
C.~M. Carloni~Calame, C.~Lunardini, G.~Montagna, O.~Nicrosini, F.~Piccinini,
  {Large angle Bhabha scattering and luminosity at flavor factories}, Nucl.
  Phys. B584 (2000) 459--479.
\newblock \href {http://arxiv.org/abs/hep-ph/0003268}
  {\path{arXiv:hep-ph/0003268}}, \href
  {http://dx.doi.org/10.1016/S0550-3213(00)00356-4}
  {\path{doi:10.1016/S0550-3213(00)00356-4}}.

\bibitem{CarloniCalame:2001ny}
C.~M. Carloni~Calame, {An improved parton shower algorithm in QED}, Phys. Lett.
  B520 (2001) 16--24.
\newblock \href {http://arxiv.org/abs/hep-ph/0103117}
  {\path{arXiv:hep-ph/0103117}}, \href
  {http://dx.doi.org/10.1016/S0370-2693(01)01108-X}
  {\path{doi:10.1016/S0370-2693(01)01108-X}}.

\bibitem{CarloniCalame:2003yt}
C.~M. Carloni~Calame, G.~Montagna, O.~Nicrosini, F.~Piccinini, {The BABAYAGA
  event generator}, Nucl. Phys. Proc. Suppl. 131 (2004) 48--55.
\newblock \href {http://arxiv.org/abs/hep-ph/0312014}
  {\path{arXiv:hep-ph/0312014}}, \href
  {http://dx.doi.org/10.1016/j.nuclphysbps.2004.02.008}
  {\path{doi:10.1016/j.nuclphysbps.2004.02.008}}.

\bibitem{Balossini:2006wc}
G.~Balossini, C.~M. Carloni~Calame, G.~Montagna, O.~Nicrosini, F.~Piccinini,
  {Matching perturbative and parton shower corrections to Bhabha process at
  flavour factories}, Nucl. Phys. B758 (2006) 227--253.
\newblock \href {http://arxiv.org/abs/hep-ph/0607181}
  {\path{arXiv:hep-ph/0607181}}, \href
  {http://dx.doi.org/10.1016/j.nuclphysb.2006.09.022}
  {\path{doi:10.1016/j.nuclphysb.2006.09.022}}.

\bibitem{Berends:1973tm}
F.~A. Berends, R.~Gastmans, {Hard photon corrections for $e^+ e^- \to \gamma
  \gamma$}, Nucl. Phys. B61 (1973) 414--428.
\newblock \href {http://dx.doi.org/10.1016/0550-3213(73)90372-6}
  {\path{doi:10.1016/0550-3213(73)90372-6}}.

\bibitem{Eidelman:1978rw}
S.~I. Eidelman, E.~A. Kuraev, {$e^+ e^-$ annihilation into two and three
  photons at high-energy}, Nucl. Phys. B143 (1978) 353--364.
\newblock \href {http://dx.doi.org/10.1016/0550-3213(78)90030-5}
  {\path{doi:10.1016/0550-3213(78)90030-5}}.

\bibitem{Berends:1980px}
F.~A. Berends, R.~Kleiss, {Distributions for electron-positron annihilation
  into two and three photons}, Nucl. Phys. B186 (1981) 22--34.
\newblock \href {http://dx.doi.org/10.1016/0550-3213(81)90090-0}
  {\path{doi:10.1016/0550-3213(81)90090-0}}.

\bibitem{CapdequiPeyranere:1978ce}
M.~Capdequi-Peyranere, G.~Grunberg, F.~M. Renard, M.~Talon, {Weak interaction
  effects in $e^+ e^- \to \gamma \gamma$}, Nucl. Phys. B149 (1979) 243--263.
\newblock \href {http://dx.doi.org/10.1016/0550-3213(79)90240-2}
  {\path{doi:10.1016/0550-3213(79)90240-2}}.

\bibitem{Bohm:1986dn}
M.~Bohm, T.~Sack, {Electroweak Radiative Corrections to $e^+ e^- \to \gamma
  \gamma$}, Z.~Phys. C33 (1986) 157--165.
\newblock \href {http://dx.doi.org/10.1007/BF01410463}
  {\path{doi:10.1007/BF01410463}}.

\bibitem{Fujimoto:1986xb}
J.~Fujimoto, M.~Igarashi, Y.~Shimizu, {Radiative Correction to $e^+ e^- \to
  \gamma \gamma$ in Electroweak Theory}, Prog. Theor. Phys. 77 (1987) 118.
\newblock \href {http://dx.doi.org/10.1143/PTP.77.118}
  {\path{doi:10.1143/PTP.77.118}}.

\bibitem{Alcaraz:2006mx}
J.~Alcaraz, et~al., {A combination of preliminary electroweak measurements and
  constraints on the standard model.~}\href
  {http://arxiv.org/abs/hep-ex/0612034} {\path{arXiv:hep-ex/0612034}}.

\bibitem{Blondel:2019vdq}
A.~Blondel, J.~Gluza, S.~Jadach, P.~Janot, T.~Riemann (Eds.), {Theory report on
  the 11th FCC-ee workshop}, 2019.
\newblock \href {http://arxiv.org/abs/1905.05078} {\path{arXiv:1905.05078}}.

\bibitem{Kuipers:2012rf}
J.~Kuipers, T.~Ueda, J.~A.~M. Vermaseren, J.~Vollinga, {FORM version 4.0},
  Comput. Phys. Commun. 184 (2013) 1453--1467.
\newblock \href {http://arxiv.org/abs/1203.6543} {\path{arXiv:1203.6543}},
  \href {http://dx.doi.org/10.1016/j.cpc.2012.12.028}
  {\path{doi:10.1016/j.cpc.2012.12.028}}.

\bibitem{Ruijl:2017dtg}
B.~Ruijl, T.~Ueda, J.~Vermaseren, {FORM version 4.2.~}\href
  {http://arxiv.org/abs/1707.06453} {\path{arXiv:1707.06453}}.

\bibitem{Montagna:1996gw}
G.~Montagna, O.~Nicrosini, F.~Piccinini, {$O(\alpha^2)$ next-to-leading
  photonic corrections to small angle Bhabha scattering in the structure
  function formalism}, Phys. Lett. B385 (1996) 348--356.
\newblock \href {http://arxiv.org/abs/hep-ph/9605252}
  {\path{arXiv:hep-ph/9605252}}, \href
  {http://dx.doi.org/10.1016/0370-2693(96)00834-9}
  {\path{doi:10.1016/0370-2693(96)00834-9}}.

\bibitem{Actis:2012qn}
S.~Actis, A.~Denner, L.~Hofer, A.~Scharf, S.~Uccirati, {Recursive generation of
  one-loop amplitudes in the Standard Model}, JHEP 04 (2013) 037.
\newblock \href {http://arxiv.org/abs/1211.6316} {\path{arXiv:1211.6316}},
  \href {http://dx.doi.org/10.1007/JHEP04(2013)037}
  {\path{doi:10.1007/JHEP04(2013)037}}.

\bibitem{Actis:2016mpe}
S.~Actis, A.~Denner, L.~Hofer, J.-N. Lang, A.~Scharf, S.~Uccirati, {RECOLA:
  REcursive Computation of One-Loop Amplitudes}, Comput. Phys. Commun. 214
  (2017) 140--173.
\newblock \href {http://arxiv.org/abs/1605.01090} {\path{arXiv:1605.01090}},
  \href {http://dx.doi.org/10.1016/j.cpc.2017.01.004}
  {\path{doi:10.1016/j.cpc.2017.01.004}}.

\bibitem{Denner:2016kdg}
A.~Denner, S.~Dittmaier, L.~Hofer, {Collier: a fortran-based Complex One-Loop
  LIbrary in Extended Regularizations}, Comput. Phys. Commun. 212 (2017)
  220--238.
\newblock \href {http://arxiv.org/abs/1604.06792} {\path{arXiv:1604.06792}},
  \href {http://dx.doi.org/10.1016/j.cpc.2016.10.013}
  {\path{doi:10.1016/j.cpc.2016.10.013}}.

\bibitem{Denner:1999gp}
A.~Denner, S.~Dittmaier, M.~Roth, D.~Wackeroth, {Predictions for all processes
  $e^+ e^- \to$~4 fermions + $\gamma$}, Nucl. Phys. B560 (1999) 33--65.
\newblock \href {http://arxiv.org/abs/hep-ph/9904472}
  {\path{arXiv:hep-ph/9904472}}, \href
  {http://dx.doi.org/10.1016/S0550-3213(99)00437-X}
  {\path{doi:10.1016/S0550-3213(99)00437-X}}.

\bibitem{Denner:2005fg}
A.~Denner, S.~Dittmaier, M.~Roth, L.~H. Wieders, {Electroweak corrections to
  charged-current $e^+ e^- \to$~4 fermion processes: Technical details and
  further results}, Nucl. Phys. B724 (2005) 247--294, [Erratum: Nucl.
  Phys.B854,504(2012)].
\newblock \href {http://arxiv.org/abs/hep-ph/0505042}
  {\path{arXiv:hep-ph/0505042}}, \href
  {http://dx.doi.org/10.1016/j.nuclphysb.2011.09.001,
  10.1016/j.nuclphysb.2005.06.033} {\path{doi:10.1016/j.nuclphysb.2011.09.001,
  10.1016/j.nuclphysb.2005.06.033}}.

\bibitem{Denner:2006ic}
A.~Denner, S.~Dittmaier, {The complex-mass scheme for perturbative calculations
  with unstable particles}, Nucl. Phys. Proc. Suppl. 160 (2006) 22--26.
\newblock \href {http://arxiv.org/abs/hep-ph/0605312}
  {\path{arXiv:hep-ph/0605312}}, \href
  {http://dx.doi.org/10.1016/j.nuclphysbps.2006.09.025}
  {\path{doi:10.1016/j.nuclphysbps.2006.09.025}}.

\bibitem{Jegerlehner:2017zsb}
F.~Jegerlehner, {Variations on photon vacuum polarization.~}\href
  {http://arxiv.org/abs/1711.06089} {\path{arXiv:1711.06089}}.

\bibitem{CarloniCalame:2011zq}
C.~Carloni~Calame, H.~Czyz, J.~Gluza, M.~Gunia, G.~Montagna, O.~Nicrosini,
  F.~Piccinini, T.~Riemann, M.~Worek, {NNLO leptonic and hadronic corrections
  to Bhabha scattering and luminosity monitoring at meson factories}, JHEP 07
  (2011) 126.
\newblock \href {http://arxiv.org/abs/1106.3178} {\path{arXiv:1106.3178}},
  \href {http://dx.doi.org/10.1007/JHEP07(2011)126}
  {\path{doi:10.1007/JHEP07(2011)126}}.

\bibitem{Jadach:1993wk}
S.~Jadach, M.~Skrzypek, B.~F.~L. Ward, {Soft pairs real and virtual infrared
  functions in QED}, Phys. Rev. D49 (1994) 1178--1182.
\newblock \href {http://dx.doi.org/10.1103/PhysRevD.49.1178}
  {\path{doi:10.1103/PhysRevD.49.1178}}.

\bibitem{Arbuzov:1995qd}
A.~B. Arbuzov, V.~S. Fadin, E.~A. Kuraev, L.~N. Lipatov, N.~P. Merenkov,
  L.~Trentadue, {Small angle electron-positron scattering with a per mille
  accuracy}, Nucl. Phys. B485 (1997) 457--502.
\newblock \href {http://arxiv.org/abs/hep-ph/9512344}
  {\path{arXiv:hep-ph/9512344}}, \href
  {http://dx.doi.org/10.1016/S0550-3213(96)00490-7}
  {\path{doi:10.1016/S0550-3213(96)00490-7}}.

\bibitem{Montagna:1998vb}
G.~Montagna, M.~Moretti, O.~Nicrosini, A.~Pallavicini, F.~Piccinini, {Light
  pair correction to Bhabha scattering at small angle}, Nucl. Phys. B547 (1999)
  39--59.
\newblock \href {http://arxiv.org/abs/hep-ph/9811436}
  {\path{arXiv:hep-ph/9811436}}, \href
  {http://dx.doi.org/10.1016/S0550-3213(99)00064-4}
  {\path{doi:10.1016/S0550-3213(99)00064-4}}.

\end{thebibliography}

\end{document}